\documentclass[10pt,letterpaper]{article}
\usepackage{opex3}
\usepackage{subeqn}
\usepackage{url}
\usepackage[normalem]{ulem}
\usepackage{cite}
\input{My_def_Opex.def}
\begin{document}

\title{Properties of nonlinear noise in long, dispersion-uncompensated fiber links }

\author{Ronen Dar,$^1$ Meir Feder,$^1$ Antonio Mecozzi,$^2$ and Mark Shtaif$^{1}$}
\address{$^1$School of Electrical Engineering, Tel Aviv University, Tel Aviv, Israel 69978\\$^2$Department of Physical and Chemical Sciences, University of L'Aquila, 67100 L'Aquila, Italy}
\email{shtaif@eng.tau.ac.il }

\begin{abstract}
We study the properties of nonlinear interference noise (NLIN) in fiber-optic communications systems with large accumulated dispersion. Our focus is on settling the discrepancy between the results of the Gaussian noise (GN) model (according to which NLIN is additive Gaussian) and a recently published time-domain analysis, which attributes  drastically different properties to the NLIN. Upon reviewing the two approaches we identify several unjustified assumptions that are key in the derivation of the GN model, and that are responsible for the discrepancy. We derive the true NLIN power and verify that the NLIN is not additive Gaussian, but rather it depends strongly on the data transmitted in the channel of interest. In addition we validate the time-domain model numerically and demonstrate the strong dependence of the NLIN on the interfering channels' modulation format.
\end{abstract}


\section{Introduction}
The modeling of nonlinear propagation in optical fibers is a key component in the efficient design of fiber-optic communications. Although computer simulations have long reached a state of maturity allowing very accurate prediction of system performance, their use is prohibitively complex in many cases of relevance, where approximate analytical models become invaluable. In a wavelength division multiplexed (WDM) environment, nonlinear propagation phenomena can be classified as either intra-channel \cite{intrachannel}, or inter-channel \cite{interchannel} effects. Intra-channel effects manifest themselves as nonlinear inter-symbol interference, which can in principle be eliminated by means of post-processing (such as back-propagation \cite{backpropagation}), or pre-distortion \cite{predistortion}. Inter-channel effects consist of cross-phase-modulation (XPM) and four-wave-mixing (FWM) between WDM channels, and in a complex network environment, {where joint processing is prohibitively complex, distortions due to inter-channel effects are random and it is customary to treat them as noise}. The chief goal of analytical models of fiber propagation is to accurately characterize this noise in terms of its statistical properties.

While early attempts of characterizing the properties of nonlinear interference noise (NLIN) in the context of fiber-communications date back to the previous millennium \cite{Petermann}, two recent analytical approaches are of particular relevance to this paper. The first approach, which  relies on analysis in the spectral domain, originated from the group of P. Poggiolini at the Politecnico di Torino \cite{Poggiolini, Carena, Poggiolini2, Poggiolini3, Carena1, Torrengo} 
and its derivation has been recently generalized by Johannisson and Karlsson \cite{Johannisson} and by Bononi and Serena \cite{Bononi}. The model generated by this approach is commonly referred to as the Gaussian noise (GN) model and its implications have already started to be addressed in a number of studies  \cite{GN_Implications1, GN_Implications2,GN_Implications3}.
The second approach has been reported by Mecozzi and Essiambre \cite{Mecozzi}, and it is based on a time-domain analysis.
The results of the latter approach \cite{Mecozzi} are distinctly different from those of the former \cite{Poggiolini, Carena, Poggiolini2, Poggiolini3, Carena1, Torrengo,Johannisson,Bononi}. Most conspicuously, in the results of \cite{Poggiolini, Carena, Poggiolini2, Poggiolini3, Carena1, Torrengo,Johannisson,Bononi}, the NLIN is treated as additive Gaussian noise and its power-spectrum is totally independent of modulation format.
Conversely, the theory of Mecozzi et al. predicts a strong dependence of the NLIN variance on the modulation format, consistently with recent experimental observations \cite{Liu}. It also predicts that in the presence of non-negligible intensity modulation a large fraction of NLIN can be characterized as phase noise. This property has a very important  practical consequence. If NLIN indeed has a large phase-noise component, as argued in \cite{Mecozzi}, then it can be canceled out easily by making use of its long temporal correlation \cite{DarArchive}, and the effective NLIN becomes much weaker than suggested by its overall variance. The consequences of this reality in terms of the predicted channel capacity have been recently studied in \cite{DarArchive,DarECOC}.

In this paper we review the essential parts of the time-domain theory of \cite{Mecozzi}, as well as those of the frequency domain GN approach. We argue that the difference between the two models results from three subtle, but very important shortcomings of the frequency domain analysis. The first is the implicit assumption that NLIN can be treated as additive noise, while ignoring its statistical dependence on the data in the channel of interest. While it is true that within the framework of a perturbation analysis NLIN can always be expressed as an additive noise term, its dependence on the channel of interest is critical. In the case of phase noise, for example, the signal of interest $s(t)$ changes into
$s(t)\exp(i\Delta\theta)$, and the noise $s(t)\exp(i\Delta\theta)-s(t)\simeq is(t)\Delta\theta$ may be uncorrelated with $s(t)$, but it is certainly not statistically independent of it. The second shortcoming of the frequency domain approach is the assumption that in the limit of large chromatic dispersion the electric field of the signal and the NLIN that accompanies it can be treated as a Gaussian processes whose  distribution is uniquely characterized in terms of its power density spectrum. The third shortcoming that we find in the GN analysis, is the claim that
non overlapping frequency components of the propagating electric field are statistically independent of each other. We show here that these components are statistically dependent in general and it is the assumption of independence that is responsible for the fact that the NLIN in \cite{Poggiolini, Carena, Poggiolini2, Poggiolini3, Carena1, Torrengo,Johannisson,Bononi} appears to be independent of modulation format. We supplement the NLIN variance obtained in the frequency domain analysis of \cite{Poggiolini} with an extra term that follows from fourth-order frequency correlations and which, as we believe, settles the discrepancy with respect to the time-domain theory of \cite{Mecozzi}.

The study contained in this paper was performed only for the case of single carrier transmission, where XPM constitutes the predominant contribution to NLIN, a fact which is confirmed by our simulations.  For this reason the analytical parts of this paper focus exclusively on XPM. Moreover, in order to isolate only the NLIN caused by inter-channel nonlinear interference, we back propagate the channel of interest so as to eliminate the distortions that are induced by SPM and chromatic dispersion.

The paper is organized as follows. In Section \ref{TDA} we review the main analytical steps of \cite{Mecozzi}, occasionally recasting them in a form that emphasizes the aspects of most relevance to this paper, and supplement them by the calculation of the autocorrelation function of the nonlinear phase-noise \cite{DarArchive}. We then review the spectral approach in Section \ref{SDA} and explain the consequences of the assumptions on Gaussianity and statistical independence that were made in \cite{Poggiolini, Carena, Johannisson, Bononi}.
In Sec. \ref{Numerics} we describe a numerical study that validates the analytical prediction of Secs. \ref{TDA} and \ref{SDA}. Section \ref{Discussion} is devoted to a summary and discussion.

\section{Time-domain analysis\label{TDA}}
We consider a channel of interest, whose central frequency is arbitrarily set to zero, and a single interfering channel whose central frequency is set to $\Omega$. Since XPM only involves two-channel interactions, the NLIN contributions of multiple WDM channels add up independently, and there is no need to conduct the initial analysis with more than a single pair. We also ignore nonlinear interactions that involve amplified spontaneous emission noise, which are negligible within the framework of a perturbation analysis such as we are conducting here. {While a second-order analysis such as in \cite{Kumar} is possible in principle, we find the first-order approach sufficiently accurate in the context of the study conducted here}.
As a starting point we express the zeroth order (i.e. linear) solution for the electric field as
\be u^{(0)}(z,t) = \sum_k a_k g^{(0)}(z,t-kT) + \sum_k b_k e^{-i \Omega t + i \frac{\beta''\Omega^2}{2}  z } g^{(0)}(z,t-kT - \beta'' \Omega z), \label{eq20}  \ee
where the superscript $^{(0)}$ throughout the equation signifies ``zeroth order".
The first sum on the right-hand-side of (\ref{eq20}) represents the channel of interest, and the second sum represents the interfering channel. The symbols $a_k$ and $b_k$ represent the data that is carried by the $k$-th symbol of the two channels, respectively, $z$ and $t$ are the space and time coordinates, $\beta''$ is the dispersion coefficient and $T$ is the symbol duration. For simplicity of notation, and without loss of generality we will assume throughout this section that $\beta''$ is negative and $\Omega$ positive. The fundamental pulse representing an individual symbol is $g^{(0)}(z,t)=\mathbf U(z)g(0,t)$, where $g(0,t)$ is the input waveform and $\mathbf U(z) = \exp\left(i\frac{1}{2}\beta''z\partial^2_t\right)$ (with $\partial_t$ denoting the time derivative operator) \cite{Comment1} is the propagation operator in the presence of chromatic dispersion. We assume that the waveform $g(0,t)$ is normalized to unit energy, whereas the actual energy of the transmitted symbols is accounted for by the coefficients $a_k$ and $b_k$. In addition it is assumed that the input waveform $g(0,t)$ is orthogonal with respect to time shifts by an integer number of symbol durations, namely $\int_{-\infty}^\infty g^*(0,t-kT)g(0,t-k'T)\df t = \delta_{k,k'}$. Owing to the unitarity of $\mathbf U(z)$ this property of orthogonality is also preserved in the linearly propagated waveform $g^{(0)}(z,t)$.

The first order correction for the field, $u^{(1)} (z, t)$, is obtained by solving the nonlinear Schr\"{o}dinger equation in which the nonlinear term is evaluated from the zeroth order approximation
\be \frac{\partial u^{(1)} (z, t)}{\partial z} = -\frac{i}{2} \beta''\partial_t^2 u^{(1)} (z, t) + i \gamma f (z) |u^{(0)} (z, t)|^2 u^{(0)} (z, t) , \label{eq30}\ee
where $\gamma$ is the nonlinearity coefficient and the function $f(z)$ accounts for the loss/gain profile along the optical link \cite{Mecozzi}. It is equal to 1 in the case of perfectly uniform distributed amplification, whereas in the case of lumped amplifiers $f(z)=\exp(-\alpha z')$, where $\alpha$ is the loss coefficient and $z'$ is the difference between the point $z$ and the position of the last amplifier that precedes it. It is assumed that only terms that contribute to the channel of interest (i.e. in the vicinity of zero frequency) are retained in the nonlinear term in (\ref{eq30}).  The solution to Eq. (\ref{eq30}) at $z=L$ is straightforward and it is given by
\be u^{(1)} (L, t) = i \gamma \int_0^L \df z \mathbf U(L-z) f (z) |u^{(0)} (z, t)|^2 u^{(0)} (z, t). \label{eq40}\ee
We now focus, without loss of generality on the detection of the zeroth data symbol $a_0$, which is obtained by passing the received field, $u(L,t)\simeq u^{(0)} (L, t)+ u^{(1)} (L, t)$, through a matched filter whose impulse response is proportional to $g^{(0)}(L,T)$. The contribution of $u^{(0)} (L, t)$ to the output of the matched filter is $a_0$ itself, whereas the contribution of $u^{(1)} (L, t)$ is the estimation error $\Delta a_0$ resulting from NLIN. It is given by
\bea \Delta a_0 = \int_{-\infty}^\infty u^{(1)}(L,t)g^{(0)^*}(L,t)\df t =
{i\gamma}  \int_0^L \df z f (z)\int_{-\infty}^\infty\df t  g^{(0)^*}(z,t)|u^{(0)} (z, t)|^2 u^{(0)} (z, t) , \label{eq50}\eea
where we have used the identity $\mathbf U(L-z)g^{(0)^*}(L,t)=g^{(0)^*}(z,t)$, which follows from the definition of the linear propagation operator. Substitution of the zeroth order field expression from Eq. (\ref{eq20}) in Eq. (\ref{eq50}) produces the result
\bea \Delta a_0 &=& i \gamma \sum_{h,k,m} \bigg(a_h a^*_k a_m S_{h,k,m} + 2 \, a_h b_k^* b_m   X_{h,k,m} \bigg). \label{eq60} \eea
where
\bea S_{h,k,m} &=& \int_0^L \df z f(z) \int \df t g^{(0)^*}(z,t) g^{(0)}(z,t-hT)g^{(0)^*}(z,t-kT)g^{(0)}(z,t-mT), \eea
is responsible for intra-channel interference effects, whereas
\bea X_{h,k,m} &=& \int_0^L \df z f(z)\int \df t g^{(0)^*}(z,t) g^{(0)}(z,t-hT)\nonumber\\
 && \times g^{(0)^*}(z,t-kT- \beta'' \Omega z) g^{(0)}(z,t-mT- \beta'' \Omega z), \label{eq80}\eea
accounts for (inter-channel) XPM induced interference. Intra-channel interference involves only symbols transmitted in the channel of interest and they need not be considered as noise. It can be reduced either by performing joint decoding of a large block of symbols, or eliminated by means of back-propagation or pre-distortion. We will hence ignore the terms proportional to $S_{h,k,m}$ in what follows and focus on the NLIN due to XPM. Notice that given the injected pulse waveform $g(0,t)$, the symbol duration $T$, the channel spacing $\Omega$ and the parameters of the fiber, the value of $X_{h,k,m}$ can be found numerically. It can be seen to reduce monotonically with the walk-off between channels, where the relevant parameter is the ratio between the group velocity difference $\beta''\Omega$ and the symbol duration $T$.

A very important feature in $X_{h,k,m}$ is that it is proportional to the overlap between {four temporally shifted waveforms. It is therefore reasonable to expect based on} Eq. (\ref{eq80}) {that the largest elements of} $X_{h,k,m}$ {are those for which} $h=0$ and $k=m$. {That is because in this situation only two temporally shifted waveforms need to overlap}. We write the contribution of these terms to $\Delta a_0$ as
\be \Delta a_{0_p} = i a_0 \left(2 \gamma  \sum_{m} |b_m|^2 X_{0,m,m}\right)=ia_0\theta, \label{eq90} \ee
where we define $\theta=2 \gamma  \sum_{m} |b_m|^2 X_{0,m,m}$. Notice that since $X_{0,m,m}$ is a real quantity according to Eq. (\ref{eq80}), $\theta$ is a real quantity as well and it represents a nonlinear phase rotation. This was the inspiration for using the sub-index $p$ (as in ``phase") in the symbol $\Delta a_{0_p}$ \cite{Comment2}. The first and second moments of $\theta$ are given by
\bea \lip\theta\rip = 2 \gamma \lip |b_0|^2\rip \sum_{m} X_{0,m,m}, \hspace{0.5cm}\mbox{and}\hspace{0.5cm}
\lip\theta^2\rip = 4 \gamma^2  \sum_{m,m'} \langle |b_m|^2 |b_{m'}|^2 \rangle X_{0,m,m} X_{0,m',m'}\nonumber\eea
and the variance of the phase rotation is
\bea \Delta{\theta}^2 = \lip\theta^2\rip-\lip\theta\rip^2=  4 \gamma^2 \left(\langle |b_0|^4 \rangle- \langle |b_0|^2\rangle ^2 \right) \sum_{m}  X_{0,m,m}^2. \label{eq120} \eea
where we have used the independence between different data symbols $\langle |b_m|^2 |b_{m'}|^2 \rangle = \langle |b_m|^2 \rangle \langle |b_{m'}|^2 \rangle (1 - \delta_{m,m'}) + \langle |b_m|^4 \rangle \delta_{m,m'}$, as well as their stationarity  $\lip |b_m|^n\rip = \lip |b_0|^n\rip$. Equation (\ref{eq120}) constitutes an extremely important  result that the phase noise grows with the variance of the square amplitude of the information symbols and that it \emph{vanishes} in the case of pure phase-modulation where $|b_0|$ is a constant (and hence $\langle |b_0|^4 \rangle- \langle |b_0|^2\rangle ^2=0$). This is a rather counter-intuitive result in view of the fact that upon propagation through a dispersive fiber, the intensity of the electric field appears to fluctuate randomly, independent of the way in which it is modulated (\cite{Poggiolini, Johannisson} and see discussion related to Fig. 2 in Sec. \ref{Numerics} of this paper).

Apart from the pure phase-noise that follows from XPM between WDM channels there are additional noise contributions involving a single pulse from the channel of interest with a pair of pulses from the interfering channel. We refer to the NLIN due to these contributions as residual NLIN, so as to distinguish it from the phase NLIN that was described earlier. In general, since residual NLIN occurs in the process of temporal overlap between three or four distinct waveforms (see Eq. (\ref{eq80}) ) its magnitude in the presence of amplitude modulation (as in 16QAM or larger QAM constellations) is expected to be notably smaller than that of phase noise, as we demonstrate numerically in section \ref{Numerics}.

A further simplification of the expression for the variance of phase-noise follows in the limit of large accumulated chromatic dispersion, which accurately characterizes the situation in most modern fiber-communications links that do not include inline dispersion compensation. In this situation the propagating waveform $g^{(0)}(z,t)$ quickly becomes proportional to its own Fourier transform \cite{DispDiff}, namely
\be g^{(0)}(z,t) \simeq \sqrt{\frac{i}{2 \pi \beta'' z}} \exp\left(-\frac{i t^2}{2 \beta'' z} \right) \widetilde g\left(0, \frac {t}{\beta'' z} \right). \label{eq130} \ee
where $\tilde g(0,\omega)=\int_{-\infty}^\infty g(0,t)\exp(i\omega t)\df t$. Equation (\ref{eq130}) simply reflects the fact that dispersion causes different frequency components of the incident signal to propagate at different velocities, so that the frequency spectrum of the injected signal is mapped into time.  In this limit the coefficients $X_{0,m,m}$ are given by
\be X_{0,m,m} = \int_{z_0}^L \df z f(z)\int \frac{\df \nu }{4 \pi^2 \beta'' z} \left| \widetilde g\left(0, \nu \right) \right|^2  \left|\widetilde g \left(0, \nu - \Omega - \frac {m T}{\beta'' z} \right) \right|^2. \label {eq120b} \ee
where we defined $\nu = t/\beta'' z$. In Eq. (\ref{eq120b}) we neglected the nonlinear distortion generated in the vicinity of the fiber input and defined $z_0 \sim T^2/|\beta''| \ll L$ as the distance after which the large dispersion approximation Eq. (\ref{eq130}) becomes valid.   Using Eq. (\ref{eq120b}) we derive an approximate analytic expression for $\Delta\theta^2$ in the case of perfectly distributed amplification.  The approximation relies on the notion that the largest overlap between the two waveforms in the integrand of (\ref{eq120b}) occurs at a position $z = z_m = - m T/\beta'' \Omega$. We replace the integral from $z_0$ to $L$ with an integral from $-\infty$ to $+\infty$ and approximate $f(z)$ with $f(z_m)$, which is set to 1 when $z_m\in[z_0,L]$ and to 0, otherwise. Physically this is equivalent to stating that all collision whose center is inside the region $[z_0,L]$ are counted as complete collisions in spite of the fact that in reality some of them (those that are centered close to the edges of the fiber) are partial. Multiplying the integrand by $z_m/z$ (which is close to unity when there is strong overlap between pulses), and changing the order of integration, we obtain
\bea X_{0,m,m} &=& \int \frac{\df \nu}{2 \pi} \left| \widetilde g\left(0, \nu \right) \right|^2 \int_{-\infty}^\infty \df z \frac{z_m f(z_m) }{2 \pi \beta'' z^2}   \left|\widetilde g \left(0, \nu - \Omega - \frac {m T}{\beta'' z} \right) \right|^2\nonumber\\
&\simeq& \left\{ \begin{array}{cc}
                          \frac{1}{\beta'' \Omega} & 0 \leq m \leq  \frac{|\beta'' \Omega|L} T \\
                          0 & \mbox{otherwise}
                        \end{array}\right.. \label{eq122}\eea
Substitution into Eq. (\ref{eq120}) yields the result
\bea \Delta{\theta}^2 = \left(\langle |b_0|^4 \rangle- \langle |b_0|^2\rangle ^2 \right) \frac{4 \gamma^2  L} {|\beta'' \Omega| T}. \label{eq160b} \eea
The simplified expression for $X_{0,m,m}$, Eq. (\ref{eq122}), also allows calculation of the temporal autocorrelation function of the phase noise $R_\theta(l) = \lip\theta_n\theta_{n+l}\rip - \lip\theta\rip^2$, where we use the notation $\theta_n$ to denote the nonlinear phase rotation induced upon the $n$-th symbol in the channel of interest. Using Eq. (\ref{eq90}) we have \cite{DarArchive,DarECOC}
\bea R_\theta(l) &=& 4\gamma^2\sum_{m}\sum_n \lip |b_m|^2|b_{n+l}|^2\rip X_{0,m,m}X^*_{0,n,n} - \lip\theta\rip^2=\Delta\theta^2 \left[1-\frac {|l|T}{|\beta''\Omega|L} \right]^+, \label{ACFtheta1}\eea
where $[a]^+=\mbox{max}\{a,0\}$. In the case of multiple WDM channels, Eq. (\ref{ACFtheta1}) generalizes to
\bea R_\theta(l) &=& \sum_s \Delta\theta^2(\Omega_s) \left[1-\frac {|l|T}{|\beta''\Omega_s|L} \right]^+,\label{ACFtheta2}\eea
where $\Omega_s$ is the frequency separation between the $s$-th WDM channel and the channel of interest and the summation is over all the interfering channels. Notice that in the limit of large accumulated dispersion, $|\beta'' \Omega_s|L/T\gg 1$, the phase noise is characterized by a very long temporal correlation. This property allows a cancelation of nonlinear phase-noise with available equalization technology \cite{PhaseEqualize1,PhaseEqualize2} and contributes to the achievement of higher information capacity \cite{DarArchive,DarECOC}. It also allows the extraction of phase noise from simulations, as we explain in Sec. \ref{Numerics}.

\section{Frequency domain analysis\label{SDA}}
In this section we review the approach adopted in \cite{Poggiolini, Carena, Poggiolini2, Poggiolini3, Carena1, Torrengo,Johannisson,Bononi} of analyzing in the frequency domain the interaction leading to NLIN and relate it to the analysis in Sec. \ref{TDA}.


Following \cite{Poggiolini}, we assume that the transmitted symbols $a_n$ and $b_n$ are periodic with period $M$, so that $a_{n+M}=a_n$, $b_{n+M}=b_n$ and the propagating field $u^{(0)}(z,t)$, which is defined in Eq. (\ref{eq20}) is periodic in time with a period $MT$. As pointed out in \cite{Poggiolini}, for large enough $M$, the assumption of periodicity is immaterial from the physical standpoint, but facilitates calculations by allowing the representation of the signal by means of discrete frequency tones,
\bea u^{(0)}(z,t) = \frac{1}{\sqrt{MT}}\left[\sum_n\nu_n(z) e^{-i\frac{2\pi}{MT}nt}+e^{-i\Omega t}\sum_n\xi_n(z) e^{-i\frac{2\pi}{MT}nt}\right].\eea
The coefficients $\nu_n$ represent the spectrum of the channel of interest at frequency $\omega = 2\pi\frac{n}{MT}$, whereas $\xi_n$ represent the spectrum of the interfering channel at $\omega = \Omega+2\pi\frac{n}{MT}$. Both $\nu_n$ and $\xi_n$ are zero mean random variables whose statistics depends on the transmitted symbols in a way on which we elaborate in what follows. The complex amplitude of the NLIN that the interfering channel imposes on the channel of interest is the sum of all the nonlinear interactions between triplets of individual frequency tones,
\bea \Delta u(t) = 2\sum_{lmn,m\neq n} \rho_{lmn}\nu_l\xi_m\xi_n^*,\label{Pogxx}\eea
where consistently with \cite{Poggiolini}, the terms $m = n$ that only contribute to a time independent phase-shift, where excluded from the summation. The factor of 2 in front of the sum in Eq. (\ref{Pogxx}) is characteristic of XPM when the nonlinearly interacting channels are co-polarized. The coefficients $\rho_{lmn}$ are  given by \cite{Carena}
\bea \rho_{lmn} &=&  \frac{\gamma}{(MT)^{3/2}}e^{-i\frac{2\pi}{MT}(l+m-n)t}\nonumber\\
&&\times\frac{1-e^{i\left(\frac{2\pi}{MT}\right)^2\beta''NL_s(m-n)(l-qM-n)}}{1-e^{i\left(\frac{2\pi}{MT}\right)^2\beta''L_s(m-n)(l-qM-n)}}\frac{1-e^{-\alpha L_s}e^{i\left(\frac{2\pi}{MT}\right)^2\beta''L_s(m-n)(l-qM-n)}}{\alpha-i\left(\frac{2\pi}{MT}\right)^2\beta''(m-n)(l-qM-n)},
\label{rho}\eea
where the WDM channel spacing is assumed to be $\Omega=q\frac{2\pi} T$, $L_s$ is the length of a single amplified span and $N$ is the overall number of amplified spans in the system. The NLIN power is given by the square average of $\Delta u(t)$
\bea \lip|\Delta u(t)|^2\rip = 4\sum_{lmnl'm'n'}
\rho_{lmn}\rho^*_{l'm'n'}\lip\nu_l\nu_{l'}^*\rip\lip\xi_m\xi_n^*\xi^*_{m'}\xi_{n'}\rip\label{Pog15},\eea
where $m\neq n$ and $m'\neq n'$ and where we have made use of the fact that $\nu_l$ and $\xi_m$ are statistically independent for all $l$ and $m$ since they correspond to different WDM channels that transmit statistically independent data. {Lack of correlation between different frequency tones implies that} $\lip\nu_l\nu_{l'}^*\rip = \lip|\nu_l|^2\rip\delta_{ll'}$, {and the assumption of true statistical independence (which is key in obtaining the results of} \cite{Poggiolini}) {implies that}
$\lip\xi_m\xi_n^*\xi^*_{m'}\xi_{n'}\rip = \lip |\xi_m|^2\rip\lip|\xi_{n}|^2\rip\delta_{mm'}\delta_{nn'}(1-\delta_{mn})$  (where the irrelevant cases with $m = n$, or $m' = n'$ were ignored for simplicity).
Equation (\ref{Pog15}) then simplifies to
\bea \lip|\Delta u|^2 \rip= 4\sum_{lmn,m\neq n}|\rho_{lmn}|^2\lip|\nu_l|^2\rip\lip|\xi_m|^2\rip\lip|\xi_n|^2\rip\label{Pog16},\eea
an expression that \emph{only} depends on the mean power spectrum of the interacting channels and is totally independent of modulation format. As we now show for the case of single carrier modulation, the above assumption of statistical independence is unjustified (even as an approximation) with most of the relevant modulation formats.

We consider a generic interfering channel as in (\ref{eq20}) $x(t) = \sum_k b_k g(t-kT)$, which is periodic as in \cite{Poggiolini} with $b_{k+M}=b_k$. The Fourier coefficients of $x(t)$ are
\bea \xi_n = \frac{1}{\sqrt{MT}}\int_0^{MT}x(t)e^{i\frac{2\pi}{MT}nt}\df t =\tilde g(\omega_n)
\frac{1}{\sqrt{MT}}\sum_{k=0}^{M-1} b_ke^{i\frac{2\pi}{M}kn}\label{Pog17},\eea
where $\tilde g(\omega)=\int g(t)\exp(i\omega t)\df t$ is the Fourier transform of $g(t)$,  $\omega_n=n\frac{2\pi}{MT}$, and the final expression on the right-hand-side follows from a straight-forward, albeit slightly cumbersome algebraic manipulation. The correlation relations between the various $\xi_n$ are obtained by averaging the product $\xi_n\xi_{n'}^*$ with respect to the transmitted data. In order to simplify the algebra we will assume Nyquist, sinc-shaped pulses $g(t) = \sinc(\pi t/T)$ in which case \cite{Poggiolini,Carena}
\bea \lip\xi_n\xi_{n'}^*\rip = \frac{|\tilde g(\omega_n)|^2}{T}\lip |b_0|^2\rip\delta_{nn'}.\label{Pog18}\eea
The restriction to Nyquist pulses ensured wide-sense stationarity for $x(t)$ and allowed avoiding the appearance of correlations between frequency tones $\xi_n$ and $\xi_{n'}$ that are separated by an integer multiple of $M$ \cite{Poggiolini}. Assuming circularly symmetric complex modulation, the  central limit theorem can be applied to Eq. (\ref{Pog17}), implying (as argued in \cite{Poggiolini,Carena}) that in the limit of large $M$, the coefficients $\xi_n$ are Gaussian distributed random variables. Yet, unlike the claim made in \cite{Poggiolini,Carena}, the fact that the coefficients $\xi_n$ are Gaussian and uncorrelated does not imply their statistical independence. That is because
the coefficients $\xi_n$ are Gaussian individually, but not jointly and hence their lack of correlation does not imply anything regarding the statistical dependence between them. In order to see the lack of joint Gaussianity note that if all $\xi_n$ were jointly Gaussian then $x(t)$ (which can be expressed as their linear combination) would have to be Gaussian as well. Therefore, unless the data-carrying symbols $b_n$ are themselves Gaussian distributed,
the Fourier coefficients $\xi_n$ cannot obey a jointly Gaussian distribution. We now write the fourth order correlation, which is obtained from Eq. (\ref{Pog17}) (again, after some algebra and for the case of Nyquist pulses)
\bea \lip\xi_m\xi_n^*\xi^*_{m'}\xi_{n'}\rip &=& \frac{\lip|b_0|^2\rip^2}{T^2}|\tilde g(\omega_m)|^2|\tilde g(\omega_{n'})|^2 \left(\delta_{m-n}\delta_{m'-n'}+\delta_{m-m'}\delta_{n-n'}\right)\nonumber\\
&&+ \frac{\lip|b_0|^4\rip-2 \lip|b_0|^2\rip^2}{MT^2}{\cal P}_{mnm'n'} \label{Pog19}\eea
where
\bea
{{\cal P}_{mnm'n'}}&=&\tilde g(\omega_m)\tilde g^*(\omega_{n})\tilde g^*(\omega_{m'})\tilde g(\omega_{n'})\nonumber\\
&&\times\left(\delta_{n-n'+m'-m-M}+\delta_{n-n'+m'-m}+\delta_{n-n'+m'-m+M}\right).\label{Pog19.5}\eea
The first term on the right-hand-side of (\ref{Pog19}) is what would follow if the coefficients $\xi_n$ where indeed statistically independent, as assumed in \cite{Poggiolini}, whereas the second term reflects the deviation from this assumption. Upon substitution into Eq. (\ref{Pog15}) we find that the noise variance  can be written as
\bea \lip|\Delta u|^2\rip&=&\lip|a_0|^2\rip\lip|b_0|^2\rip^2\chi_1 + \lip|a_0|^2\rip\left(\lip|b_0|^4\rip-2 \lip|b_0|^2\rip^2\right)\chi_2\label{PogFinal}\eea
where
\bea \chi_1 &=& \frac{4}{T^3}\sum_{lmn,m\neq n}|\tilde g(\omega_l)|^2|\tilde g(\omega_m)|^2|\tilde g(\omega_n)|^2|\rho_{lmn}|^2, \label{Pog24}\\
\chi_2 &=& \frac{4}{MT^3}\sum_{lmnm'n'}|\tilde g(\omega_l)|^2{{\cal P}_{mnm'n'}}\rho_{lmn}\rho_{lm'n'}^*\label{Pog25}
\eea
where terms with $m=n$ or $m'=n'$ are excluded from the summation. The first term on the right-hand-side of (\ref{PogFinal}) is due to second-order correlations between the frequency tones and we will refer to it as the second-order noise (SON). This term coincides with the result of \cite{Poggiolini,Johannisson} (and can be obtained by substituting Eq. (\ref{Pog18}) in Eq. (\ref{Pog16})). The second term is absent in the calculations of \cite{Poggiolini,Johannisson} and since it results from fourth order correlations between the frequency tones we will refer to it as fourth-order noise (FON). Consistently, we will refer to $\chi_1$ and $\chi_2$ as the SON and FON coefficients, respectively.
Due to the delta functions in the definition of ${{\cal P}_{mnm'n'}}$ in Eq. (\ref{Pog19.5}), the number of free indices in the summation in Eq. (\ref{Pog25}) is four (e.g. $l,m,n,m'$, in which case $n'$ is determined by the other four and given by $n'=n+n'-m-M$, $n'=n+n'-m$, or $n'=n+n'-m+ M$). Since every free index runs over $O(M)$ values, the factor of $1/M$ in the expression for the FON coefficient $\chi_2$ is countered by $M$ more summations than in Eq. (\ref{Pog24}) and hence $\chi_1$ and $\chi_2$ are of similar order of magnitude. Moreover, as we demonstrate numerically in Sec. \ref{Numerics}, in the limit of distributed amplification the coefficients $\chi_2$ is almost identical to $\chi_1$ and they become practically indistinguishable when  the frequency separation between the interfering channels grows (see Fig. \ref{chi12}).
Interestingly, in the special case of purely Gaussian modulation, when the symbols $b_k$ are circularly symmetric complex Gaussian variables, $\lip|b_0|^4\rip-2 \lip|b_0|^2\rip^2=0$ and the FON vanishes, in which case the NLIN spectrum found in \cite{Poggiolini} is exact. Consistently, we remind that this is also the only case in which $x(t)$ is truly Gaussian distributed and the lack of correlation between different frequency tones indeed implies their statistical independence.

The last point that we address in this section is the assumption of Gaussianity in the context of NLIN in the limit of high chromatic dispersion. The argument against this assumption is similar to the argument made in the context of Gaussianity in the frequency domain. That is because in the limit of large dispersion, the signal frequency spectrum is simply mapped to the time domain. Therefore, the field becomes Gaussian point-wise, but it does not form a Gaussian process. It is in fact a general principle that a linear unitary time independent operation, such as chromatic dispersion, cannot transform a non-Gaussian process into a Gaussian one. In the absence of joint Gaussianity between all of the field samples, the power density spectrum does not sufficiently characterize the nature of NLIN.

\section{Numerical validation\label{Numerics}}
In order to validate the analytical results of the previous section, a set of simulations, all based on the  standard split-step Fourier transform method, was performed. In order to demonstrate the principle and to be able to test the phase-noise variance predicted in Eq. (\ref{eq160b}) we perform all simulations for the case of perfectly distributed gain, namely where the loss coefficient $\alpha$ is set to 0. The simulations are performed for a 500 km system over a standard single mode fiber, whose dispersion coefficient is $\beta''=21$ ps$^2/$km and whose nonlinearity coefficient is given by $\gamma=1.3$ W$^{-1}$km$^{-1}$. As we are only interested in characterizing the NLIN, we did not include ASE noise in any of the simulations.
In all our simulations the {symbol-rate} was 100 Gb/s, similarly to \cite{Essiambre}, and the channel spacing was set to 102 GHz. Nyquist pulses of a perfectly square optical spectrum (of 100 GHz width) were assumed. The number of simulated symbols in each run was 8192 and up to 500 runs (each with independent and random data symbols) were performed with each set of system parameters, so as to accumulate sufficient statistics. The data symbols of the various channels were generated independently of each other using Matlab's random number generator whose periodicity is much larger than the collective number of symbols produced in our simulations. Use of very long sequences in every run is critical in such simulations so as to achieve acceptable accuracy in view of the long correlation time of NLIN, as well as to avoid artifacts related to the periodicity of the signals that is imposed by the use of the discrete Fourier transform.
In all system simulations that we present, the number of WDM channels was five, with the central channel being the channel of interest. At the receiver the channel of interest was isolated with a matched optical filter and back-propagated so as to eliminate the effects of SPM and chromatic dispersion.
\begin{figure}[t!]
\centering\includegraphics[width=0.7\columnwidth]{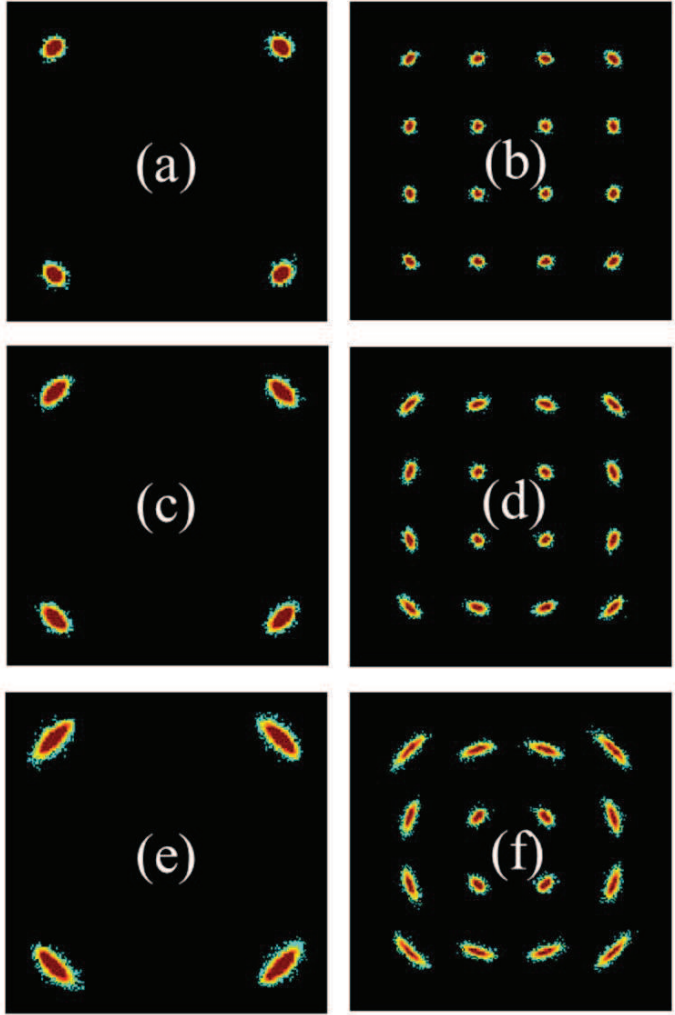}
\caption{Received constellations (after compensating for the average nonlinear phase-rotation) of the channel of interest after 500 km of fiber. The per-channel power was -2dBm. \color{black}The channel of interest is QPSK-modulated in the left column and 16-QAM modulated in the right column. In the top panel (Figs. (a) and (b)) the modulation of the interfering channels is QPSK. In the middle panel (Figs. (c) and (d)) the modulation of the interfering channels is 16-QAM, and in the bottom panel (Figs. (e) and (f)) the modulation of the interfering channel is Gaussian. The dominance of phase noise is evident in the middle and bottom panels, whereas in the top panel phase-noise is negligible.}\label{Constellations}
\end{figure}
\subsection{Modulation format dependence}
In order to demonstrate the dependence of NLIN on the modulation format we plot in Fig. \ref{Constellations} the received signal constellations in six different cases. The figures in the left column represent the case in which the \textit{channel of interest} undergoes QPSK modulation, whereas the right column refers to the case in which the modulation of the channel of interest is 16-QAM. The figures in the top panel correspond to the case in which the \textit{interfering channels} are QPSK modulated, whereas the figures in the middle panel were produced with 16-QAM modulated interferers. The bottom two figures were produced in the case where the symbols of the interfering channels where Gaussian modulated.
In the top panel, where the interfering channels undergo pure phase modulation, the NLIN is almost circular, albeit a small amount of phase-noise can still be observed. This small phase-noise is due to coefficients $X_{0,k,m}$ ($k\neq m$) that were neglected in Sec. \ref{TDA} \cite{Comment2}.
In the center and bottom panels, where the intensity of the interfering channels is modulated, the phase-noise nature of NLIN is very evident, and it is largest in the case of Gaussian modulation.

\begin{figure}[t!]
\centering\includegraphics[width=0.7\columnwidth]{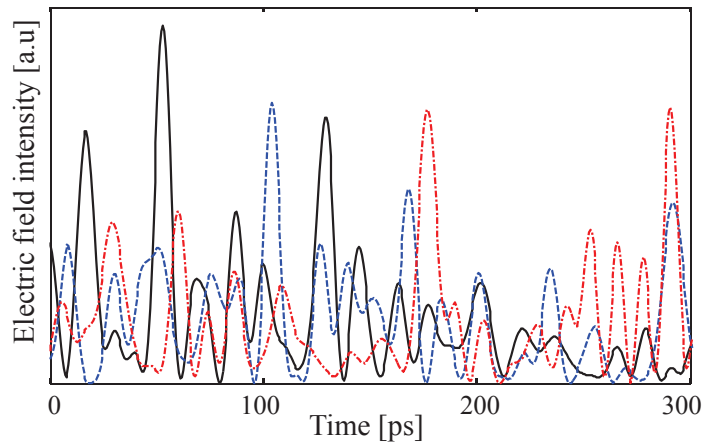}
\caption{The electric field intensities of a single channel operating with Nyquist sinc-shaped pulses at a baud-rate of 100 GHz after being dispersed by 8500 ps/nm/km (equivalent to 500 km in standard fiber). The solid (black), dashed (blue), and dash-dotted (red) curves correspond to QPSK, 16-QAM and Gaussian modulation, respectively. In spite of the apparent similarity between the dispersed waveforms as demonstrated in this figure, the NLIN strongly depends on the modulation format.}\label{Intensities}
\end{figure}

The modulation format dependence that is predicted in \cite{Mecozzi} and summarized in Sec. \ref{TDA} is of somewhat subtle origin and is fairly counter-intuitive. As was argued correctly in \cite{Poggiolini,Johannisson,Bononi} the electric field of the strongly dispersed signal appears fairly random independently of the modulation format as can be seen in Fig. \ref{Intensities}. Moreover, as noted earlier, the point-wise distribution of the field is indeed Gaussian. Nonetheless the types of NLIN produced by the various modulations are very different as can be clearly seen in Fig. \ref{Constellations}.

We note that the phase-noise nature of NLIN was not evident in the simulation results reported in \cite{Carena}. While the difference between the results cannot be determined unambiguously based on the simulation details provided in \cite{Carena}, it may result from certain differences in the simulated system. Most importantly, the simulations in \cite{Carena} do not eliminate intra-channel effects through back-propagation, as we do here, but use adaptive equalization, which may leave some of the intra-channel interference uncompensated. Furthermore, it is possible that the phase-noise that we report (which is characterized by a very long temporal correlation) is inadvertently eliminated in the process of adaptive equalization. Additionally, some of the discrepancy could result from the fact that the system simulated in \cite{Carena} assumed lumped amplification, as opposed to distributed amplification that we assumed here. It is possible that these differences explain the agreement between the simulations reported in \cite{Carena} and the analytical results of the GN model.
\begin{figure}[t!]
\centering\includegraphics[width=0.7\columnwidth]{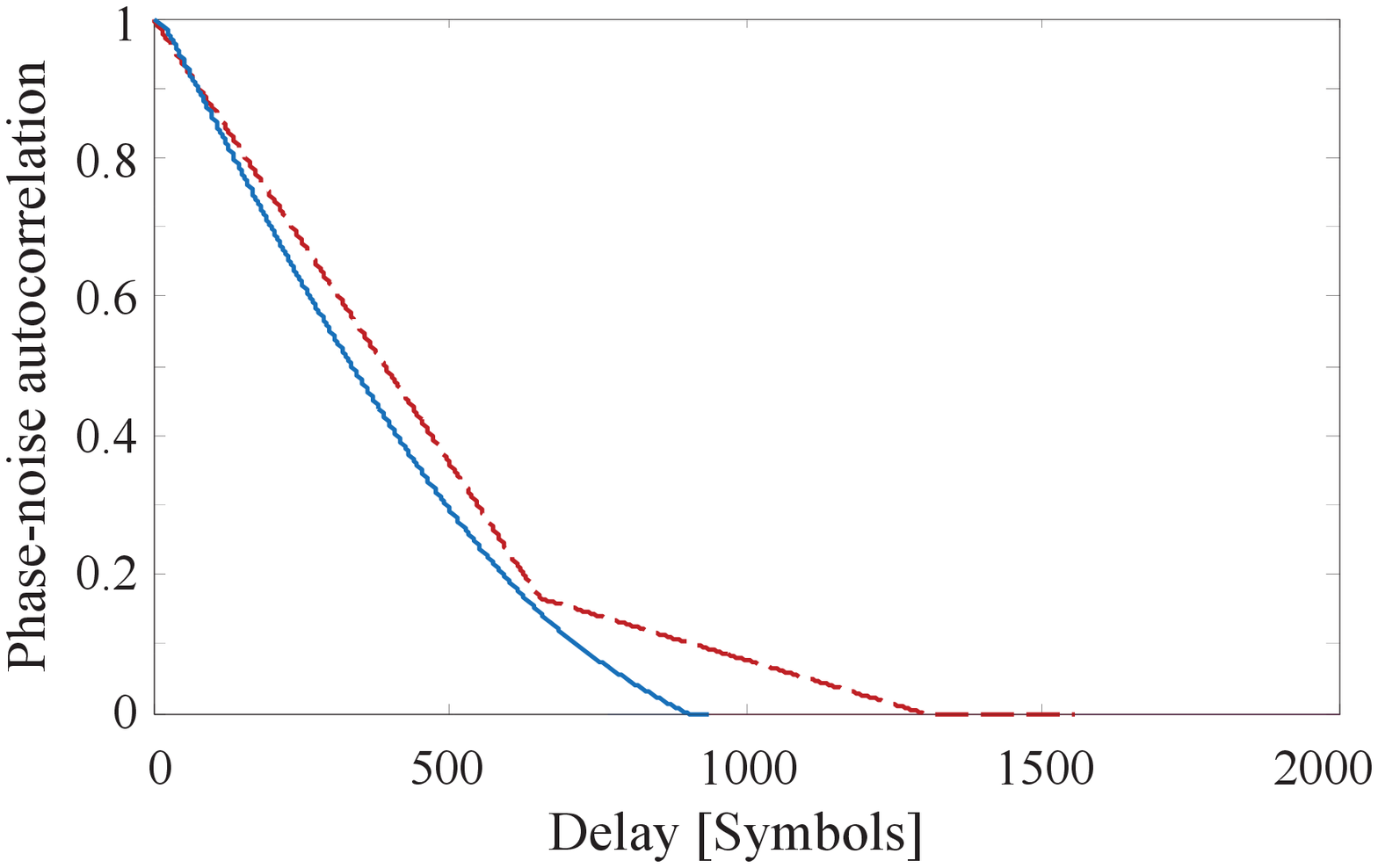}
\caption{The phase-noise autocorrelation function $R_\theta(l)$ of the phase-noise Eq. (\ref{ACFtheta2}) (dashed-red) and as obtained from the simulations (solid blue) for -6dBm per-channel average power \cite{accuracy}.}\label{ACFtheta}
\end{figure}
\begin{figure}[t!]
\centering\includegraphics[width=0.7\columnwidth]{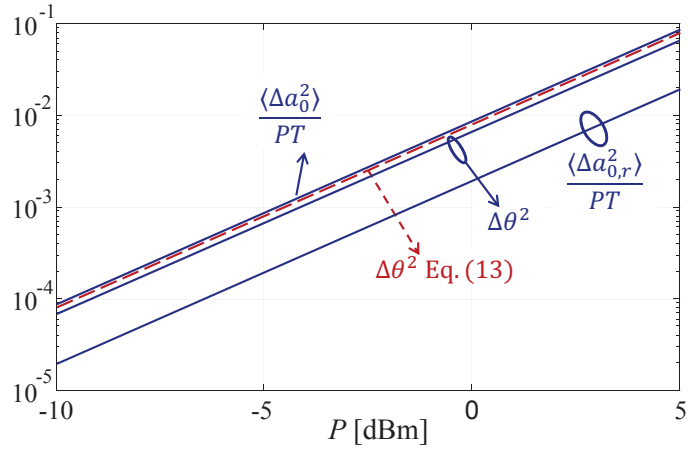}
\caption{The complete NLIN variance $\lip\Delta a_0^2\rip$ normalized to the average symbol energy (top solid curve), the phase noise $\Delta\theta^2$ as obtained from the simulations (center solid curve) and the variance of the residual noise $\lip\Delta a_r^2\rip$ normalized to the average symbol energy \cite{accuracy}. The dashed curve (red) shows the analytical result for $\Delta\theta^2$, Eq. (\ref{eq160b}). {It is within 20\% of the numerically obtained $\Delta\theta^2$}. }\label{Variances}
\end{figure}
\subsection{The variance of phase-noise and assessment of the residual NLIN}
{In this section we validate the analytical expression for the phase-noise variance in Eq. (\ref{eq160b}), and assess the residual noise. We remind that the residual noise is the part of the NLIN that does not manifest itself as phase-noise and hence remains after phase-noise cancelation. To this end, we define a procedure for extracting the phase noise from the results of the simulations.}
Denoting by $r_n$ the $n$-th sample of the received signal (in the channel of interest and after back propagation and matched filtering) we have $r_n = a_n\exp(i\theta_n)+\Delta a_{n,r}$, where  $\Delta a_{n,r}$ is the residual noise.
We  extract $\theta_n$ through a least-squares procedure by performing a sliding average of the quantity $a_n^*r_n$ over a moving window of $N=50$ adjacent symbols. We then normalize the absolute value of the averaged quantity to 1, so as to ensure that we are only extracting phase noise. The residual noise $\Delta a_{n,r}$ is evaluated by subtracting $a_n\exp(i\hat\theta_n)$ (with $\hat\theta_n$ being the estimated phase) from the received sample $r_n$.
The width of the sliding window needs to be narrow enough relative to the correlation time of $\theta_n$, but broad enough to ensure meaningful statistics. Using this procedure we computed the autocorrelation function of the nonlinear phase $\theta$, which is plotted in Fig. \ref{ACFtheta} together with the analytical expression (\ref{ACFtheta2}). The agreement between the analytical and numerical autocorrelation functions is self evident. Notice that over a block of 50 symbols the autocorrelation of $\theta_n$ drops only by 6\% relative to its maximal value, thereby justifying the choice of $N=50$ for the moving average window. Further considerations in optimizing the window-size can be found in \cite{DarArchive}.

Figure \ref{Variances} shows the normalized overall NLIN variance $\lip\Delta a_0^2\rip/PT$, the phase-noise variance $\Delta\theta^2$, and the normalized variance of the residual noise $\lip\Delta a_{0,r}^2\rip/PT$, where $P$ is the average power in each of the interfering channels. The analytical expression for the phase-noise variance Eq. (\ref{eq160b}) is also plotted by the dashed red curve. All the curves in Fig. \ref{Variances} were obtained in the case of Gaussian modulation of the data-symbols. The accuracy of the analytical result is self evident, as is the clear dominance (that was predicted in \cite{Mecozzi}) of the phase-noise component of NLIN.
\begin{figure}[t!]
\centering\includegraphics[width=0.9\columnwidth]{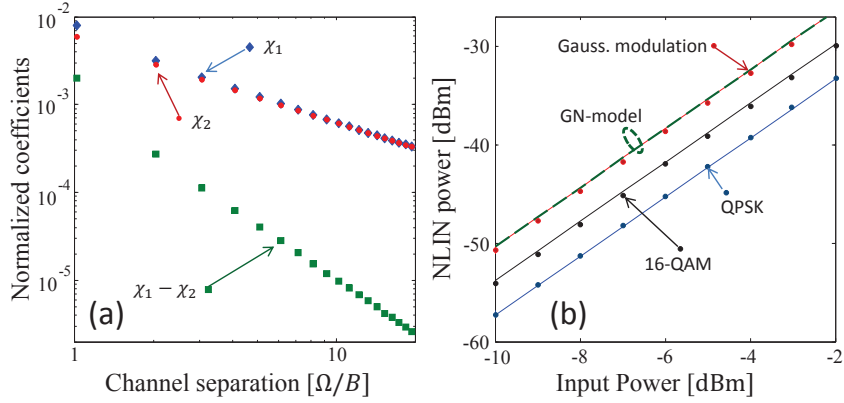}
\caption{(a) The SON coefficient $\chi_1$ (blue diamonds) and the FON coefficient $\chi_2$ (red circles) as a function of the spacing between channels. The coefficients in the figure are normalized by $\gamma^2L^2/T^3$ and hence they are unit-less. The green squares show $\chi_1-\chi_2$. (b) The NLIN power versus the average power per-channel for QPSK, 16-QAM and Gaussian modulation. The symbols show the results of a split-step-simulation performed with the same parameters as in Fig. 1. The solid lines represent Eq. (\ref{Our_var}), whereas the dashed green line shows the prediction of the GN model Eq. (\ref{GN_var}). It is correct only with Gaussian modulation, but severely overestimates the actual noise in other formats.}\label{chi12}
\end{figure}
\subsection{The difference with respect to the NLIN power predicted by the GN model}
In order to assess the error in the estimation of the NLIN power by the GN model, we compute the NLIN power, as it is predicted by the GN model and as it is predicted by the theory in Sec. \ref{SDA}. In the case where $P$ is the average power used in each of the channels, these quantities are specified by Eq. (\ref{PogFinal}) and given by
\bea \lip|\Delta u|^2\rip_{\mathrm{GN}} &=& P^3\sum_s\chi_1(\Omega_s)\label{GN_var}\\
\lip|\Delta u|^2\rip_{\mathrm{Full}} &=& P^3\sum_s[\chi_1(\Omega_s) -\chi_2(\Omega_s)]+P^3\left(\frac{\lip |b_0|^4\rip}{\lip|b_0|^2\rip^2}-1\right)\sum_s\chi_2(\Omega_s), \label{Our_var} \eea
where the summation index $s$ runs over all neighboring channels (which are spectrally separated by $\Omega_s$ from the channel of interest).
The SON coefficient $\chi_1$ and the FON coefficient $\chi_2$ are plotted in Fig. \ref{chi12}a as a function of the frequency separation between the interacting channels, where the blue diamonds are used to represent $\chi_1$ and the red circles represent $\chi_2$. The two coefficients are seen to be very similar to each other so that the difference between them, which is illustrated by the green squares, is significantly smaller than the coefficients themselves. The Monte-Carlo integration method \cite{MonteCraloI} was deployed in order to compute the sums in  Eqs. (\ref{Pog24}) and (\ref{Pog25}) in the limit of $M\to\infty$ with the estimation error being always lower than 3\%.

In Fig. \ref{chi12}b we show the NLIN power in our simulated 5-channel system in the cases of QPSK, 16-QAM, and Gaussian modulation. The this solid curves show the theoretical result $\lip|\Delta u|^2\rip_{\mathrm{Full}}$ of Eq. (\ref{Our_var}) and the circles represent the variance obtained in the full split-step simulation. The dashed green line represents the prediction of the GN model $\lip|\Delta u|^2\rip_{\mathrm{GN}}$,  Eq. (\ref{GN_var}), which is correct only for Gaussian modulation. In the case of QPSK the actual NLIN power is lower by approximately 6.5dB than the prediction of the GN model. Since the NLIN powers in Fig. \ref{chi12}b  include the contribution of phase-noise, the relation to the error-rate is not straightforward.

\section{Discussion\label{Discussion}}
Having reviewed the essential parts of the time domain model and the frequency domain GN model, we have pointed out that the difference between the models results from three unjustified assumptions in the frequency domain approach. The assumption that NLIN can be described as an additive noise term that is statistically independent on the signal, the assumption that in the large dispersion limit the electric field of the signal and the noise forms a Gaussian process that is uniquely characterized in terms of its spectrum,  and the claim of statistical independence between non-overlapping tones in the spectrum of the interfering signal. We have shown that by correctly accounting for fourth-order correlations in the signals' spectrum an extra term --- the FON --- arises. The FON (which can be positive, or negative depending on the modulation format) needs to be added to the noise power obtained in the GN model (the SON) in order to obtain the correct overall NLIN. The inclusion of the FON recovers the dependence of the NLIN power on modulation format, a property that is absent from the existing GN model and reconciles between the frequency domain and the time domain theories. We stress that the current GN model of \cite{Poggiolini, Carena, Poggiolini2, Poggiolini3, Carena1, Torrengo,Johannisson,Bononi} which does not contain the FON term, cannot be considered a valid approximation, since with standard modulation formats (e.g QPSK, 16-QAM), the magnitude of the FON is comparable to that of the SON, which is the quantity calculated in \cite{Poggiolini,Carena}.
The numerical validation of the theoretical results has been performed in the case of a five-channel WDM system with idealized distributed amplification. In this case the FON term was almost identical to the SON term, implying that the error in the NLIN power predicted by the GN model is very significant.

While the study presented in this paper focused on the single polarization case, the effect of polarization multiplexing can be be anticipated by considering the relevant factors. The SON part of the NLIN variance changes in the presence of polarization multiplexing by a factor of 16/27 \cite{Carena}, whereas it can be shown that the FON part changes by 40/81. The small difference between these factors has practically no effect on the conclusions made in this paper regarding the importance of accounting for FON. The numerical study of polarization multiplexed transmission, as well as the effects of lumped amplifications and the many other practical system parameters, is beyond the scope of this work and will be addressed in the future.

Finally, we note that when treating the  NLIN as an additive, signal-independent noise process, its bandwidth appears to be comparable to that of the signal itself. Thus, one cannot take advantage of the fact that phase noise that dominates the variance of NLIN in many cases of interest is very narrow-band as we have shown here (see Eq. (\ref{ACFtheta2}) and Fig. \ref{ACFtheta}). The importance of this property of NLIN is immense as it allows cancelation of the phase-noise part of NLIN by means of available equalization technology \cite{PhaseEqualize1,PhaseEqualize2}, such that the residual NLIN (whose variance is much smaller than that of the NLIN as a whole) determines system performance. The system consequences of this reality have been addressed in \cite{DarECOC}.

\section*{Acknowledgement}
Financial support from the Israel Science Foundation (grant 737/12) is gratefully acknowledged. Ronen Dar would like to acknowledge the support of the Adams Fellowship Program of the Israel Academy of Sciences and Humanities, and the Yitzhak and Chaya Weinstein Research Institute for Signal Processing. The authors acknowledge useful comments by P.J Winzer and A. Bononi.


\begin{thebibliography}{99}
\bibitem{Comment1} We have assumed a fixed position independent dispersion parameter $\beta''$. The generalization to position dependent dispersion is straightforward, but we avoid it as it considerably complicates the notation.
\bibitem{Comment2} Although terms involving $X_{0,k,m}$ with $m\neq k$ also contribute to phase noise, they are much smaller than the terms proportional to $X_{0,m,m}$ (particularly in the limit of large dispersion) and we omit them from the analysis presented in this paper.
\end{thebibliography}
\end{document}